\documentclass[10pt,conference]{IEEEtran}
\IEEEoverridecommandlockouts
\usepackage{cite}
\usepackage{amsmath,amssymb,amsfonts}
\usepackage{graphicx}
\usepackage{textcomp}
\usepackage{xcolor}

\usepackage[table]{xcolor}
\usepackage{colortbl}
\usepackage{algpseudocode}
\usepackage{times}
\usepackage{soul}
\usepackage{url}
\usepackage{subcaption}
\usepackage[hidelinks]{hyperref}
\usepackage[utf8]{inputenc}
\usepackage[]{caption}
\usepackage{amsmath}
\usepackage{amsthm}
\usepackage{hyperref}

\usepackage{cleveref}
\usepackage{graphicx}
\graphicspath{{./figs/}{../}}
\usepackage{booktabs}
\usepackage{algorithm}
\usepackage[switch]{lineno}
\usepackage{threeparttable}
\newcommand{\minisection}[1]{\vspace{.06in}\noindent{\textbf{#1}}.}
\usepackage{array}
\usepackage{listings}
\usepackage{xcolor}
\usepackage{booktabs}
\usepackage{multirow}
\usepackage{balance}
\definecolor{red}{RGB}{255, 0, 0}         
\definecolor{orangeRed}{RGB}{255, 69, 0}  
\definecolor{orangeYellow}{RGB}{255, 165, 0} 
\definecolor{yellowGreen}{RGB}{154, 205, 50} 
\definecolor{lightGreen}{RGB}{144, 238, 144} 
\definecolor{darkGreen}{RGB}{0, 128, 0}   

\def\BibTeX{{\rm B\kern-.05em{\sc i\kern-.025em b}\kern-.08em
    T\kern-.1667em\lower.7ex\hbox{E}\kern-.125emX}}

\begin{document}

\title{Think with Self-Decoupling and Self-Verification: Automated RTL Design with Backtrack-ToT

\thanks{\textdagger~ Corresponding Author
\\This paper has been published in the 29th Design, Automation and Test in Europe Conference (DATE 2026), April 20-22, 2026,
Verona, Italy.}
}

\author{
\IEEEauthorblockN{
Zhiteng Chao\textsuperscript{1},
Yonghao Wang\textsuperscript{1\textdagger},
Xinyu Zhang\textsuperscript{1,2},
Jiaxin Zhou\textsuperscript{3}, 
Tenghui Hua\textsuperscript{1,2},
Husheng Han\textsuperscript{1},
Tianmeng Yang\textsuperscript{2},
\\
Jianan Mu\textsuperscript{1\textdagger},
Bei Yu\textsuperscript{5},
Rui Zhang\textsuperscript{1},
Jing Ye\textsuperscript{1,2}
and
Huawei Li\textsuperscript{1,2}}
\IEEEauthorblockA{\textsuperscript{1}State Key Lab of Processors, Institute of Computing Technology, CAS, Beijing, China}
\IEEEauthorblockA{\textsuperscript{2}University of Chinese Academy of Sciences, Beijing, China}
\IEEEauthorblockA{\textsuperscript{3}Beijing Normal University, China; \textsuperscript{4}Peking University, China; \textsuperscript{5}Chinese University of Hong Kong, Hongkong}

\IEEEauthorblockA{\{chaozhiteng, wangyonghao22s, mujianan, lihuawei\}@ict.ac.cn}
}


\maketitle
\thispagestyle{plain}
\pagestyle{plain}

\begin{abstract}


Large language models (LLMs) hold promise for automating integrated circuit (IC) engineering using register transfer level (RTL) hardware description languages (HDLs) like Verilog. However, challenges remain in ensuring the quality of Verilog generation. Complex designs often fail in a single generation due to the lack of targeted decoupling strategies, and evaluating the correctness of decoupled sub-tasks remains difficult. While the chain-of-thought (CoT) method is commonly used to improve LLM reasoning, it has been largely ineffective in automating IC design workflows, requiring manual intervention. The key issue is controlling CoT reasoning direction and step granularity, which do not align with expert RTL design knowledge. This paper introduces VeriBToT, a specialized LLM reasoning paradigm for automated Verilog generation. By integrating Top-down and design-for-verification (DFV) approaches, VeriBToT achieves self-decoupling and self-verification of intermediate steps, constructing a Backtrack Tree of Thought with formal operators. Compared to traditional CoT paradigms, our approach enhances Verilog generation while optimizing token costs through flexible modularity, hierarchy, and reusability.
\end{abstract}


\section{Introduction}

Digital circuits lie at the core of modern computing systems, yet their design remains a demanding process. Engineers rely on register transfer level (RTL) hardware description languages (HDLs) to translate functional requirements into logic gate combinations, which are then verified and mapped into transistor-level designs for fabrication. Although the downstream tasks of verification and transistor mapping have been largely automated, the initial writing of HDL code still demands significant human effort. This deficiency represents a pressing bottleneck in chip design workflows, inflating costs and undermining efficiency~\cite{chen2024dawn}.
\begin{figure}[tb!]
    \centering
    \begin{minipage}{0.6\linewidth}
        \centering
        \includegraphics[width=\linewidth]{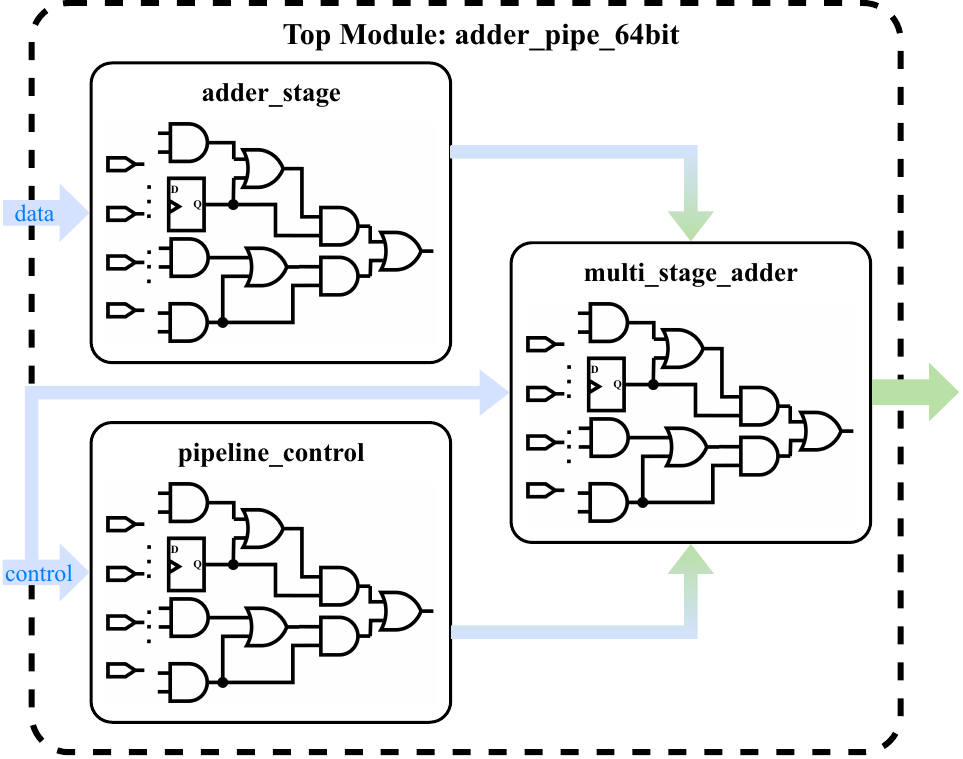}
        \subcaption{}
        \label{figa}
    \end{minipage}
    \begin{minipage}{0.36\linewidth}
        \centering
        \includegraphics[width=\linewidth]{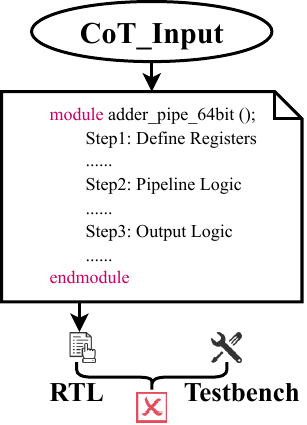}
        \subcaption{}
        \label{figb}
    \end{minipage}
    \begin{minipage}{1\linewidth}
        \centering
        \includegraphics[width=\linewidth]{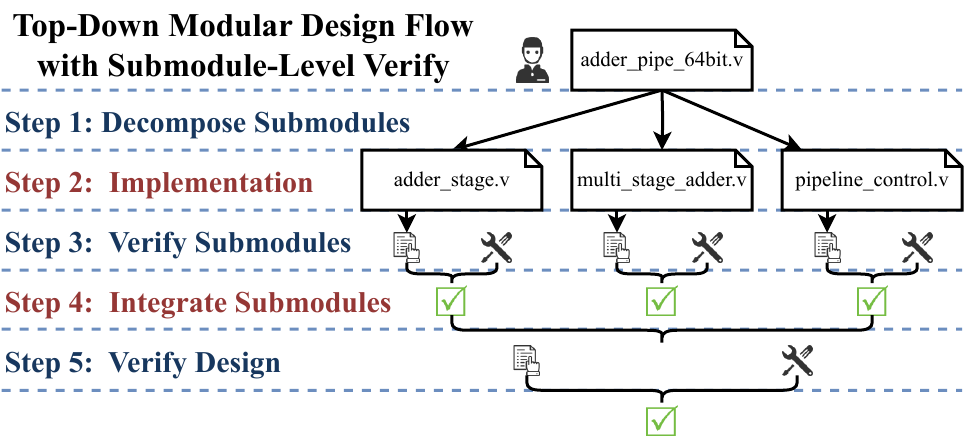}
        \subcaption{}
        \label{figc}
    \end{minipage}
    \caption{The modular concurrency in circuit design is illustrated in (a), (b) presents common scenarios for the CoT paradigm in generating an RTL module's Verilog code step by step, (c) depicts the human circuit design paradigm.}
    \label{fig:cot_exam}
\end{figure}

Recent advances in artificial intelligence, notably in large language models (LLMs)~\cite{achiam2023gpt}, present an encouraging avenue for mitigating this challenge. In fields like natural language processing (NLP), LLMs have achieved remarkable success in software~\cite{hou2023large} and hardware code generation~\cite{chang2024data,ChatCPU,RTLCoder,delorenzo2024make,liu2023verilogeval,Thakur,lu2024rtllm,zhao2024codev,VeriGen}. This triumph underscores their potential to automate and streamline complex workflows.

A pivotal factor enabling the success of LLMs in software code generation is the ``chain-of-thought'' (CoT) reasoning paradigm~\cite{wei2022chain,Structured,le2023codechain}, which decomposes intricate problems into smaller, sequential steps. By enhancing the accuracy and logical consistency of generated outputs, CoT significantly advances automated code generation.
However, directly applying CoT methods to the domain of automated RTL design has yielded limited results as sequence reasoning process is vulnerable and hard to capture hierarchical dependencies of complex submodules.

We conducted a task-specific analysis to better understand the challenges of naive CoT in automated RTL design.
Circuit design is fundamentally different from describing sequential functional behaviors in programming languages like C or Python. As illustrated in Fig.~\ref{figa}, all functional modules in RTL design are implemented as concrete hardware circuits and operate in parallel to achieve the desired functionality. This parallelism introduces numerous interdependencies between modules. For LLMs, generating RTL requires considering a significantly broader range of contextual information across multiple RTL modules, as well as achieving precise cycle-level design coupling. Consequently, reasoning and decision-making in RTL design are considerably more complex than in software code generation, which typically involves understanding only a limited range of and coarse grained preceding contexts. When naive CoT is applied to enhance LLMs for generating RTL as shown in Fig.~\ref{figb}, it is often observed that the generated results are uncontrollable, producing unnecessary and meaningless content, without improving the generation capability of LLMs.

The RTL design process by which human engineers write code is illustrated in Fig.~\ref{figc} and exhibits two key characteristics from an engineering design perspective. Firstly, engineers consider factors such as circuit complexity and adopt a Top-down approach~\cite{kaeslin2014top} to partition the design into modules. This ensures that each module remains highly cohesive internally while maintaining low coupling with other modules, thereby reducing both the difficulty of developing individual modules and the complexity of integrating them. Secondly, during the coding process, engineers verify each module’s functionality before merging them, iterating layer by layer~\cite{dfv}. This approach significantly lowers the overall debugging complexity. Together, these two practices facilitate efficient iteration and integration in modular design.

In this paper, we introduce VeriBToT, a novel Backtrack-Tree-of-Thought framework for modular RTL design, inspired by human IC design methods. Our approach divides the LLM reasoning process into two cognitive pathways: self-decoupling for partition refinement, and self-verification for ensuring the correctness of generated code. To implement these pathways, we design five prompt generation operators based on human IC design interactions, integrated into a Backtrack Tree of Thought. Experiments on open-source benchmarks show that VeriBToT enhances code correctness compared to existing CoT methods. The main contributions of this work are as follows:
\begin{enumerate}
\item We propose a novel Backtrack-Tree-of-Thought framework that integrates IC design methodologies (e.g., Top-down and DFV), achieving refined control over the thought process for LLM automatic RTL generation.
\item By leveraging the modular and hierarchical principles of the Top-down methodology, we decouple the ToT thought step nodes into corresponding RTL submodules within the IC design. The leaf nodes, representing frontier thought steps, are empowered to self-decouple their leaf-module designs into lower-level submodules.
\item By leveraging Top-down and DFV methodologies, we decouple common self-refinement operations found in traditional prompt engineering~\cite{madaan2024self},  achieving modular verification and design-verification co-iteration in thought process, granting the intermediate thought steps the capability of self-verification.
\item Extensive experimental results show that the proposed VeriBToT significantly improves code correctness, with a significant improvement on the benchmark compared to existing CoT methods.
\end{enumerate}

\section{Motivation and Related Work}

\begin{figure}[tb]
    \center
    \includegraphics[width=8.5cm]{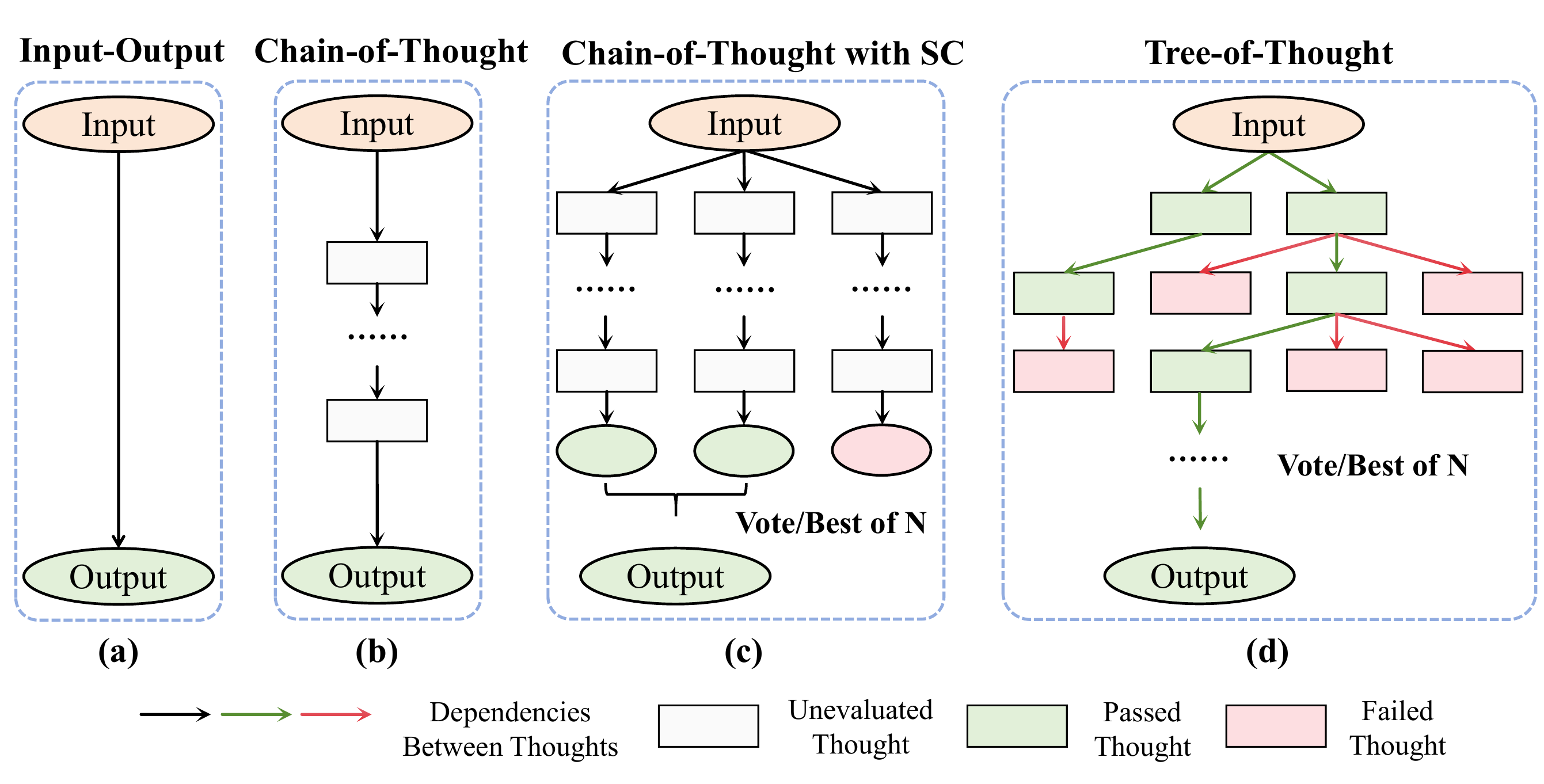}
    \caption{
        The comparison between traditional CoT paradigms.
    }
    \label{fig1_trans}
\end{figure}

CoT-based prompt engineering~\cite{wei2022chain} is used to improve reasoning and problem-solving performance for LLMs.
The core idea is to explicitly guide the model through a multi-step reasoning by breaking down a complex task into smaller, sequential steps.
This approach mimics human thought processes. 
Numerous variants have been developed around CoT, such as CoT-SC~\cite{wang2022self}, tree of thoughts (ToT)~\cite{yao2024tree,long2023large},
and graph of thoughts (GoT)~\cite{besta2024graph}. The comparison of different CoT reasoning paradigms is illustrated in the Fig.~\ref{fig1_trans}. The traditional CoT method offers some degree of assistance in the field of code generation~\cite{le2023codechain,huang2023codecot,wang2024intervenor,ni2024tree,VToT}.

However, there are obstacles to enhancing RTL generation capabilities by directly applying any traditional CoT. Current CoT approaches lack the modular design and verification practices essential for hardware code development. These methods follow a linear, unidirectional flow, where large functional modules are decomposed without reconsidering the reasonableness of the partitioning. This often leads to overly complex modules or excessive inter-module interactions, making it difficult for LLMs to generate correct and integrable code.
Moreover, CoT lacks single step verification, relying instead on functional validation after completing the entire code. This approach results in unclear feedback, making it hard to pinpoint and address specific issues. The lack of fine-grained control in current CoT methods also can lead to a significant increase in token consumption. As demonstrated by traditional ToT in Fig.~\ref{fig1_trans}d, both the tree depth and the number of child nodes for each node are hyperparameters, and setting these values too large can incur substantial computational costs. Consequently, the absence of modular thinking and iterative verification are key challenges that limit CoT's effectiveness in RTL generation.

\begin{figure}[tb]
    \center
    \includegraphics[width=9cm]{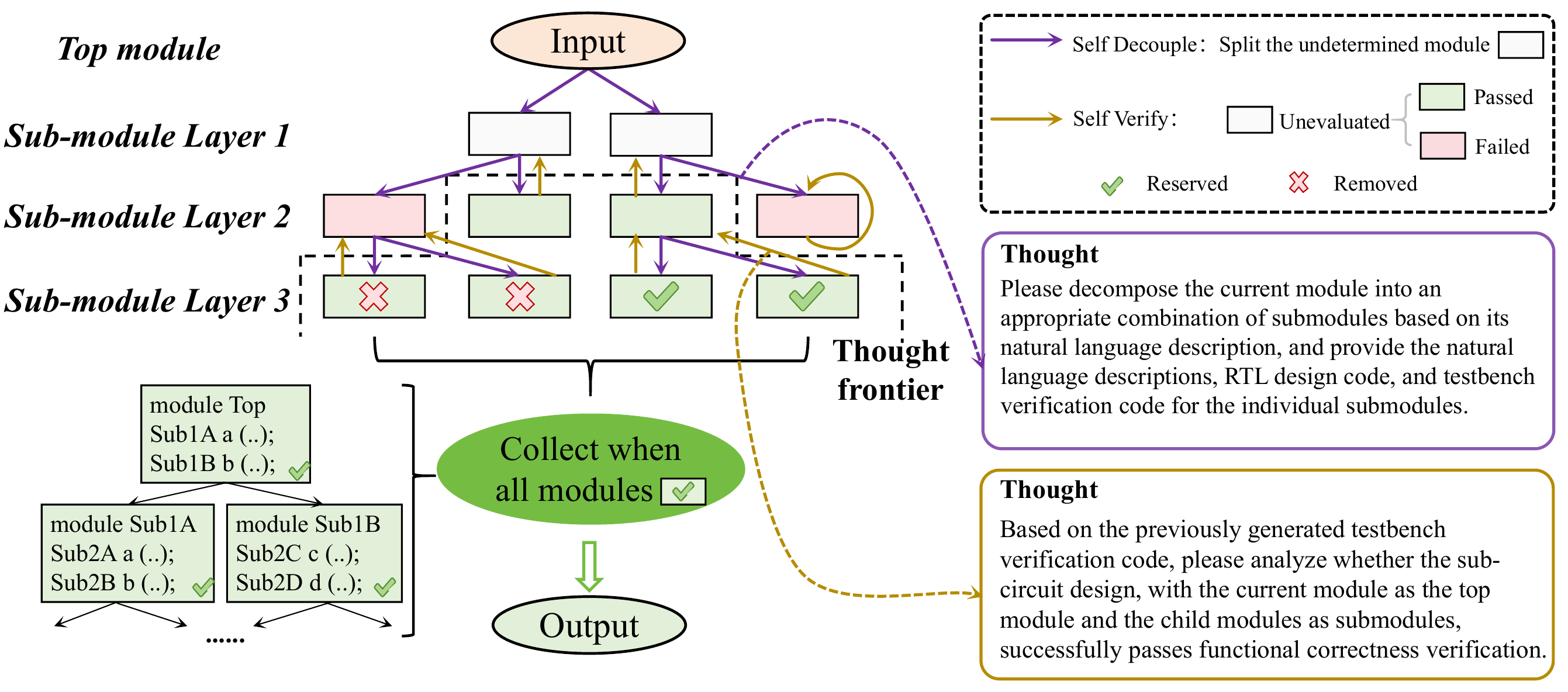}
    \caption{
        The motivation of VeriBToT: each node in the diagram represents a thought step, which is limited to a module within the hierarchical RTL design (including natural language descriptions, RTL design Verilog, and testbench Verilog). 
    }
    \label{fig1_new}
\end{figure}

Inspired by human-centric IC design methodologies (Top-down, DFV), we introduce a dedicated ToT paradigm, VeriBToT, for automatic RTL generation by LLMs as shown in Fig.~\ref{fig1_new}. 
The Top-down~\cite{kaeslin2014top} approach is a hierarchical design methodology that starts with system-level requirements and progressively refines the design to lower abstraction levels, promoting early validation, design reuse, and team collaboration.
Design-for-verification (DFV)~\cite{dfv} enhances the verifiability of designs by incorporating verification-friendly features early in the development process, aiming to reduce verification complexity and improve efficiency through modularization and better test planning.
The process begins at inputting the natural language description of top-level module design. Then, self-decoupling thought prompt is designed for hierarchical sub-module decomposition to minimize inter-module dependencies, thereby enabling each sub-module to be designed and verified independently without affecting the others. 
Each thought step generates a complete design flow for a sub-module, including natural language descriptions, RTL design Verilog, and testbench Verilog. For instance, a top-level module~\emph{Top} may be decomposed into multi-layer lower level sub-modules (e.g., ~\emph{Sub1A} could contain ~\emph{Sub2A} and ~\emph{Sub2B}, and ~\emph{Sub1B} could contain ~\emph{Sub2C} and ~\emph{Sub2D}). When to invoke the self-decoupling thought requires the LLM to self-assess the complexity of the current design.

The self-verify thought prompt is designed to verify the correctness of the RTL code generated for each sub-module. Based on the already generated testbench, LLMs assess whether the RTL code for the sub-design rooted at the current node is functionally correct. Starting from the leaf node, the process iterates upward. If the LLMs determine that the implementation at the current node is incorrect, the existing sub-design is discarded and rewritten.

\section{The Proposed VeriBToT Framework}




\subsection{Overview}
We designed VeriBToT, a specialized LLM ToT reasoning paradigm for automated RTL design, featuring thought-step self-decoupling and self-verification capabilities.
As shown in Fig.~\ref{fig1_new}, the core of VeriBToT lies in how to formally define these two cognitive pathways. Inspired by the practical IC design process, we define five specific operations that structure the entire reasoning process.

\subsection{Operator Design}

We have designed five operators to perform VeriBToT. These operators structure the reasoning process into a tree of thought, and the interactions between the operators can be abstracted as collaboration among different design personnel in an IC company design workflow, with the entire process adhering to the principles of Top-down and DFV design.


\begin{figure*}[tb!]
    \center
    \includegraphics[width=18cm]{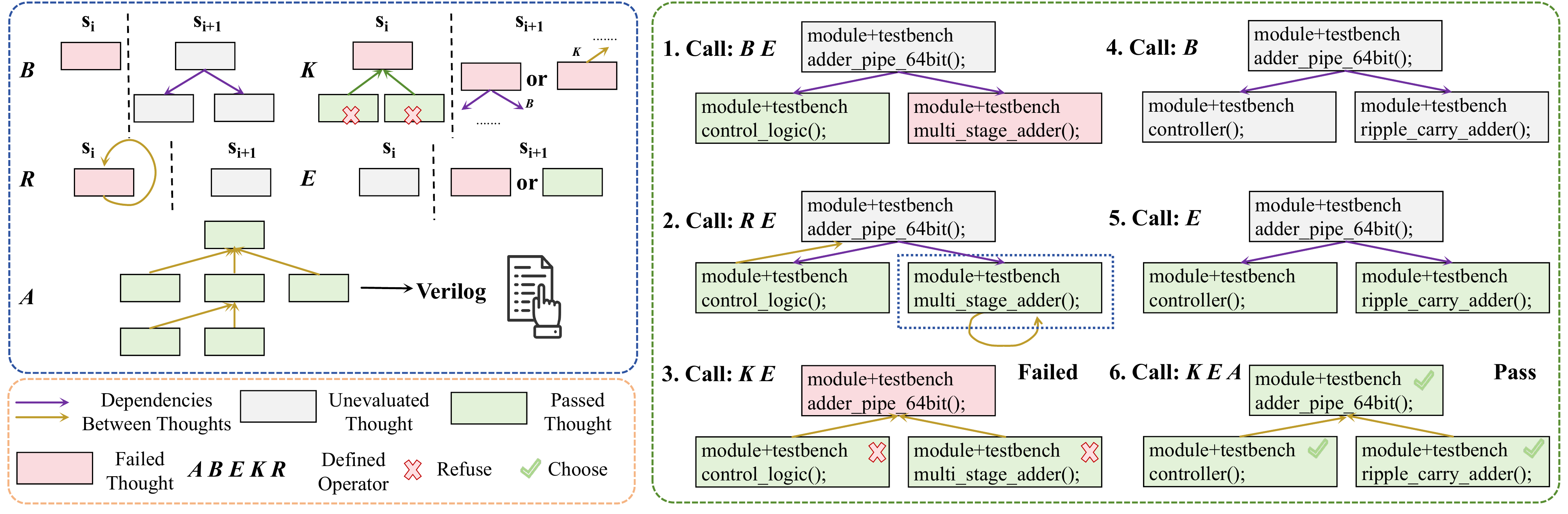}
    \caption{The left side of the figure illustrates a schematic representation of how pre-defined operators control the thought process. The right side depicts how a 64-bit adder is generated through the thought process of VeriBToT, achieved through the combination of these operators; to differentiate among the five operators we have defined, we employ the following abbreviations:~\textit{Branch Generator} is represented by \(B\),~\textit{Node Evaluator} by \(E\),~\textit{Node Rethinker} by \(R\),~\textit{Backtrack Executor} by \(K\), and~\textit{Code Aggregator} by \(A\).}
    \label{fig2}
\end{figure*}
\begin{enumerate}
\item \textit{Branch Generator}: Under a correct Top-down design module decomposition plan created by the product manager, RTL designers and testers will break down a complex design into appropriate submodules and separately design the RTL and testbench for each submodule.

\item \textit{Node Evaluator}: Testers follow the module decomposition plan designed by the product manager to create separate test plans for different levels of modules according to their specifications. Once the RTL designs for all submodules under the current subtree are completed, the tester uses the pre-designed testbench to verify whether the RTL functionality of the current subsystem is correct.

\item \textit{Node Rethinker}: When a single bottom-level module designed by an engineer fails through the testbench, the tester requests the design engineer to immediately redesign the RTL of this module.

\item \textit{Backtrack Executor}: An engineer responsible for a middleware module composed of lower-level modules discovers during the actual development process that the product manager's module decomposition plan contains errors, making it impossible to produce middleware that meets overall design requirements. The engineer requests that the product manager revise the module decomposition plan to determine whether the next step should involve redesigning based on the current middleware or modifying the design of the layer above. 
\item \textit{Code Aggregator}: The product manager ultimately assembles all the validated hierarchical modules into a complete design for approval.

\end{enumerate}

By implementing these operators, each intermediate thought step can achieve self-decoupling and self-validation, enhancing the overall design ability of LLMs.

\subsection{Reasoning Process}

We define several notions used in this paper based on concepts from previous methods to facilitate the formalization of the problem. 
We use~\textbf{lowercase letters} \( x, y, z \) to represent~\textbf{language sequences}. Here, \( x \) denotes the original natural language description of the RTL design, serving as the input for the entire problem; \( y \) represents the final answer to the problem, including the complete Verilog code as the output; and \( z_1, \ldots, z_n \) denotes a~\textbf{chain of thoughts} that bridges \( x \) and \( y \), where each \( z_i \) is a intermediate thought step that guides the process of RTL generation.
In the~\textbf{ToT paradigm}, LLMs are permitted to explore multiple reasoning paths across the nodes of a tree, where an intermediate~\textbf{state} in the tree is represented as \( s = [x, z_1, \ldots, z_i] \), a~\textbf{node of thought} is represented as \( n \) sampled from the current state \( s \), and a collection of all states for ToT is denoted by \( S =[s_1, \ldots, s_n]\).

The formal representation of the five operators and a complete example of the VeriBToT reasoning process are illustrated in Fig.~\ref{fig2}. 
To formalize the discussion of the exploration process of VeriBToT, we use~\textbf{uppercase letters}: \( L\) to denote natural language, \(D\) to denote RTL design, and \(V\) to denote testbench for verification. From the perspective of the content of language sequences, each intermediate thought step \( z_i \) can be split into three parts: the RTL design language description~\( z^{L}_i \), the Verilog code for the current module RTL design~\( z^{D}_i \), and the testbench for the current design~\( z^{V}_i \).
The entire process can be broadly categorized into the following key components, organized according to the depth-first search (DFS) traversal method of the Backtrack Tree algorithm:

\definecolor{mycolor1}{rgb}{0.4, 0.8, 0.1}
\definecolor{mycolor3}{rgb}{0.8, 0.5, 0}
\definecolor{mycolor32}{rgb}{1, 0.8, 0.1}
\definecolor{mycolor4}{rgb}{1, 0.75, 0.7}
\definecolor{mycolor5}{rgb}{0.9, 0.3, 0.3}
\begin{table*}[h]
\caption{Reasoning paradigm evaluation on the full benchmark about functional \emph{Pass@1} and \emph{Pass@5}.}
        \label{tab:full}
    \centering

    \begin{tabular}{@{}lp{1.2cm}p{1.2cm}p{1.2cm}p{1.2cm}p{1.2cm}p{1.2cm}p{1.2cm}p{1.2cm}p{1.2cm}p{1.2cm}@{}}
        \toprule
        \multirow{2}{1.5 cm}{Reasoning Paradigm} & \multicolumn{4}{c}{VerilogEval-Human} & \multicolumn{4}{c}{RTLLM} 
        \\ \cmidrule(lr){2-5} \cmidrule(lr){6-9}
                           & \multicolumn{2}{c}{\textbf{DeepSeek}} & \multicolumn{2}{c}{\textbf{ChatGPT-4}} & \multicolumn{2}{c}{\textbf{DeepSeek}} & \multicolumn{2}{c}{\textbf{ChatGPT-4}} \\
        \cmidrule(lr){2-3} \cmidrule(lr){4-5} \cmidrule(lr){6-7} \cmidrule(lr){8-9}
                           & \textbf{\emph{pass@1}} & \textbf{\emph{pass@5}} & \textbf{\emph{pass@1}} & \textbf{\emph{pass@5}} & \textbf{\emph{pass@1}} & \textbf{\emph{pass@5}} & \textbf{\emph{pass@1}} & \textbf{\emph{pass@5}} \\ \midrule
        IO     & 0.27 & 0.36 & \cellcolor{lightGreen}0.34 & \cellcolor{lightGreen}0.45 & \cellcolor{lightGreen}0.34 & \cellcolor{lightGreen}0.46 & \cellcolor{mycolor1}0.42 & \cellcolor{mycolor1}0.56 \\
        CoT    & 0.29 & 0.37 & \cellcolor{lightGreen}0.33 & \cellcolor{mycolor1}0.47 & \cellcolor{lightGreen}0.30 & \cellcolor{lightGreen}0.40 & \cellcolor{lightGreen}0.38 & \cellcolor{mycolor1}0.50\\
        CoT-SC & 0.27 & 0.37 & \cellcolor{mycolor1}0.38 & \cellcolor{mycolor1}0.50 & \cellcolor{lightGreen}0.24 & \cellcolor{lightGreen}0.38 & \cellcolor{mycolor1}0.40 & \cellcolor{lightGreen}0.46 \\
        ToT    & 0.26 & 0.39 & \cellcolor{lightGreen}0.35 & \cellcolor{mycolor1}0.48 & \cellcolor{lightGreen}0.28 & \cellcolor{lightGreen}0.48 & \cellcolor{lightGreen}0.38 & \cellcolor{lightGreen}0.48 \\
        VeriBToT   & \cellcolor{lightGreen}0.32 & \cellcolor{lightGreen}0.44 & \cellcolor{mycolor1}0.43 & \cellcolor{mycolor1}0.58 & \cellcolor{mycolor1}0.42 & \cellcolor{mycolor1}0.56 & \cellcolor{mycolor1}0.48 & \cellcolor{mycolor1}0.62  \\ 
        VeriBToT-   & 0.29 &0.36 & \cellcolor{lightGreen}0.36 & \cellcolor{mycolor1}0.52 & \cellcolor{lightGreen}0.26 & \cellcolor{lightGreen}0.44 & \cellcolor{mycolor1}0.41 & \cellcolor{mycolor1}0.57  \\\midrule
        \multirow{2}{1.5 cm}{Domain Specific} 
                           & \multicolumn{2}{c}{\textbf{Thakur}} & \multicolumn{2}{c}{\textbf{RTLCoder}} & \multicolumn{2}{c}{\textbf{Thakur}} & \multicolumn{2}{c}{\textbf{RTLCoder}} \\
        \cmidrule(lr){2-3} \cmidrule(lr){4-5} \cmidrule(lr){6-7} \cmidrule(lr){8-9}
                           & \textbf{\emph{pass@1}} & \textbf{\emph{pass@5}} & \textbf{\emph{pass@1}} & \textbf{\emph{pass@5}} & \textbf{\emph{pass@1}} & \textbf{\emph{pass@5}} & \textbf{\emph{pass@1}} & \textbf{\emph{pass@5}} \\ \midrule
        IO & 30.3 & 43.9 & 36.7 & 45.5 & 14.9 & 24.1 & 39.8 & 48.3\\
        \bottomrule
    \end{tabular}
            \begin{tablenotes}
    \item[1] Domain specific models lack the ability to follow instructions based on different reasoning paradigms; the light green indicates that the current result surpasses one domain-specific fine-tuned model, while the dark green indicates that it surpasses two.
    \end{tablenotes} 
        
\end{table*}

\definecolor{mycolor1}{rgb}{0.4, 0.8, 0.1}
\definecolor{mycolor3}{rgb}{0.8, 0.5, 0}
\definecolor{mycolor32}{rgb}{1, 0.8, 0.1}
\definecolor{mycolor4}{rgb}{1, 0.75, 0.7}
\definecolor{mycolor5}{rgb}{0.9, 0.3, 0.3}
\begin{table*}
\caption{The performance of the Deepseek-Coder-V2 in generating Verilog for different hard cases under various reasoning paradigms is measured by the number of successful attempts out of five trials (\#\emph{pass@5}), assessing syntax correctness (Syn.) and functional correctness (Fun.); the average token consumption (Tok., the unit is thousand, \emph{k}) is used to evaluate the resource expenditure of the reasoning paradigms, \textbf{Bold values} indicate results from our method that are the best.}
    \label{tab:Deepseek}
    \centering
     \footnotesize
    \setlength{\tabcolsep}{5pt} 
        \resizebox{0.88\linewidth}{!}{
    \begin{tabular}{|p{2.5cm}|p{0.6cm}|p{0.6cm}|p{0.6cm}|p{0.6cm}|p{0.6cm}|p{0.6cm}|p{0.6cm}|p{0.6cm}|p{0.6cm}|p{0.6cm}|p{0.6cm}|p{0.6cm}|p{0.6cm}|p{0.6cm}|p{0.6cm}|}
        \hline
        \multirow{2}{*}{Design} & \multicolumn{3}{c|}{IO} & \multicolumn{3}{c|}{CoT} & \multicolumn{3}{c|}{CoT-SC} & \multicolumn{3}{c|}{ToT} & \multicolumn{3}{c|}{VeriBToT}\\
        \cline{2-16}
        & Syn. & Fun. & Tok. &  Syn. & Fun. & Tok. &  Syn. & Fun. & Tok. & Syn. & Fun. & Tok.&  Syn. & Fun. & Tok. \\
        \hline
        adder\_16bit & \cellcolor{mycolor3!60}3 & \cellcolor{mycolor3!60}3 & 1.11 & \cellcolor{mycolor1}5 & \cellcolor{mycolor3!60}3 & 1.35 & \cellcolor{lightGreen}4 & \cellcolor{mycolor3!60}3 & 2.62 & \cellcolor{lightGreen}4 & \cellcolor{mycolor3!60}3 & 1.39 & \cellcolor{mycolor1}\textbf{5}  & \cellcolor{mycolor1}\textbf{5} & 1.88\\
        \hline
        adder\_pipe\_64bit & \cellcolor{mycolor1}5 & \cellcolor{mycolor4}1 & 0.67 & \cellcolor{mycolor1}5 & \cellcolor{mycolor4}1 & 0.82 & \cellcolor{lightGreen}4 & \cellcolor{mycolor32!60}2 & 1.47 & \cellcolor{mycolor1}5 & \cellcolor{mycolor3!60}3 & 1.94 & \cellcolor{mycolor1}\textbf{5}  & \cellcolor{mycolor32!60}2 & 2.34\\
        \hline
        multi\_pipe\_8bit & \cellcolor{mycolor5}0 & \cellcolor{mycolor5}0 & 0.94 & \cellcolor{mycolor5}0 & \cellcolor{mycolor5}0 & 1.26 & \cellcolor{mycolor5}0 & \cellcolor{mycolor5}0 & 2.03 & \cellcolor{mycolor3!60}3 & \cellcolor{mycolor5}0 & 1.86 & \cellcolor{lightGreen}\textbf{4} & \cellcolor{mycolor5}0 & 1.41\\
        \hline
        multi\_16bit & \cellcolor{mycolor4}1 & \cellcolor{mycolor4}1 & 1.1 & \cellcolor{mycolor3!60}3 & \cellcolor{mycolor5}0 & 1.02 & \cellcolor{mycolor4}1 & \cellcolor{mycolor5}0 & 1.92 & \cellcolor{lightGreen}4 & \cellcolor{mycolor32!60}2 & 1.45 & \cellcolor{lightGreen}\textbf{4}  & \cellcolor{mycolor4}1 & 1.58\\
        \hline
        barrel\_shifter & \cellcolor{mycolor1}5 & \cellcolor{mycolor5}0 & 0.62 & \cellcolor{mycolor1}5 & \cellcolor{mycolor5}0 & 0.67 & \cellcolor{lightGreen}4 & \cellcolor{mycolor5}0 & 1.65 & \cellcolor{lightGreen}4 & \cellcolor{mycolor4}1 & 1.41 & \cellcolor{mycolor1}\textbf{5}  & \cellcolor{mycolor4}\textbf{1}  & 2.12\\
        \hline
        width\_8to16 & \cellcolor{mycolor1}5 & \cellcolor{mycolor5}0 & 0.53 & \cellcolor{mycolor32!60}2 & \cellcolor{mycolor4}1 & 1.01 & \cellcolor{mycolor32!60}2 & \cellcolor{mycolor4}1 & 0.97 & \cellcolor{mycolor1}5 & \cellcolor{mycolor4}1 & 0.99 & \cellcolor{lightGreen}4 & \cellcolor{mycolor3!60}\textbf{3} & 1.62\\
        \hline
        calendar & \cellcolor{mycolor1}5 & \cellcolor{mycolor5}0 & 0.52 & \cellcolor{mycolor1}5 & \cellcolor{mycolor5}0 & 0.98 & \cellcolor{mycolor32!60}2 & \cellcolor{mycolor5}0 & 1.38 & \cellcolor{mycolor1}5 & \cellcolor{mycolor5}0 & 1.1 & \cellcolor{mycolor1}\textbf{5}  & \cellcolor{mycolor1}\textbf{5} & 1.45\\
        \hline
        freq\_div & \cellcolor{mycolor1}5 & \cellcolor{mycolor5}0 & 0.74 & \cellcolor{mycolor1}5 & \cellcolor{mycolor4}1 & 1.16 & \cellcolor{mycolor1}5 & \cellcolor{mycolor32!60}2 & 1.08 & \cellcolor{mycolor3!60}3 & \cellcolor{mycolor5}0 & 1.78 & \cellcolor{mycolor1}\textbf{5}  & \cellcolor{lightGreen}\textbf{4} & 1.44\\
        \hline
        dff8r      & \cellcolor{mycolor3!60}3 & \cellcolor{mycolor5}0 & 0.43 & \cellcolor{lightGreen}4 & \cellcolor{lightGreen}4 &0.65 & \cellcolor{mycolor5}0 & \cellcolor{mycolor5}0 & 1.41& \cellcolor{mycolor3!60}3 & \cellcolor{mycolor5}0 & 1.23 & \cellcolor{mycolor1}5 & \cellcolor{mycolor1}5&1.21\\
        \hline
        fsm2s      & \cellcolor{mycolor5}0 & \cellcolor{mycolor5}0 &0.64 & \cellcolor{mycolor5}0 & \cellcolor{mycolor5}0 & 0.78& \cellcolor{mycolor5}0 & \cellcolor{mycolor5}0 &1.52 & \cellcolor{mycolor5}0 & \cellcolor{mycolor5}0 & 1.10 & \cellcolor{mycolor32!60}2 &  \cellcolor{mycolor4}1& 1.82\\
        \hline
        fsm3comb   & \cellcolor{mycolor5}0 & \cellcolor{mycolor5}0 & 0.63 & \cellcolor{mycolor4}1 & \cellcolor{mycolor4}1 & 0.90 & \cellcolor{mycolor4}1 & \cellcolor{mycolor4}1 & 1.68 & \cellcolor{mycolor5}0 & \cellcolor{mycolor5}0 & 1.34 & \cellcolor{mycolor1}5 & \cellcolor{mycolor3!60}3& 2.56\\
        \hline
        gates4     & \cellcolor{mycolor1}5 & \cellcolor{mycolor4}1 & 0.42 & \cellcolor{mycolor1}5 & \cellcolor{lightGreen}4 &0.65 & \cellcolor{lightGreen}4 & \cellcolor{mycolor3!60}3 & 1.36 & \cellcolor{mycolor1}5 & \cellcolor{lightGreen}4 & 1.24 & \cellcolor{mycolor1}5 & \cellcolor{mycolor1}5& 1.49\\
        \hline
        popcount3  & \cellcolor{mycolor5}0 & \cellcolor{mycolor5}0 & 0.57 & \cellcolor{mycolor3!60}3 & \cellcolor{mycolor4}1 & 0.76 & \cellcolor{mycolor5}0 & \cellcolor{mycolor5}0 & 1.25 & \cellcolor{mycolor4}1 & \cellcolor{mycolor5}0 & 1.77 & \cellcolor{mycolor32!60}2 & \cellcolor{mycolor32!60}2& 1.77\\
        \hline
        m2014\_q4i	&\cellcolor{mycolor5}0	&\cellcolor{mycolor5}0	&0.13&\cellcolor{mycolor5}0	&\cellcolor{mycolor5}0	&0.35&\cellcolor{mycolor5}0	&\cellcolor{mycolor5}0	&0.82& \cellcolor{mycolor3!60}3	&\cellcolor{mycolor32!60}2	&1.06&\cellcolor{lightGreen}4	&\cellcolor{lightGreen}4& 0.42\\
        \hline
        ringer	&\cellcolor{mycolor5}0	&\cellcolor{mycolor5}0	&0.39&\cellcolor{mycolor5}0	&\cellcolor{mycolor5}0	&0.71&\cellcolor{mycolor5}0	&\cellcolor{mycolor5}0	&0.91&\cellcolor{mycolor5}0	&\cellcolor{mycolor5}0	&1.06&\cellcolor{lightGreen}4	&\cellcolor{mycolor3!60}3&2.53\\
        \hline
        timer	&\cellcolor{mycolor5}0	&\cellcolor{mycolor5}0	&0.47&\cellcolor{mycolor32!60}2	&\cellcolor{mycolor32!60}2	&0.68&\cellcolor{mycolor5}0	&\cellcolor{mycolor5}0	&1.29&\cellcolor{mycolor32!60}2	&\cellcolor{mycolor4}1	&1.15&\cellcolor{lightGreen}4	&\cellcolor{mycolor3!60}3&1.77\\
        \hline
    \end{tabular}
    }
    
\end{table*}

\minisection{Tree Initialization} 
The process begins with a root node \( s_0 = x \), \( n_0 = x \). The input sequence \( x = [x^L, x^V] \) represents a natural language description of the overall RTL design and the testbench used for final functionality testing.

\minisection{Branching} 
For a newly generated leaf node \( n = [n^L, n^D, n^V]\), VeriBToT calls~\textit{Node Evaluator} for self-verifying \(n^D\) by \(n^V\), both of which are generated at the same step for \(n\). If the LLM determines that the validation is successful, it proceeds to the next node. If the validation fails again, the node faces two choices. If the LLM determines that the current design is still relatively complex, it decides to invoke the \textit{Branch Generator}. The LLM, using a ``branching prompt,'' decomposes the current leaf node’s design into several new leaf nodes, each of which describes the complete design requirements \(n'^{L}\), implementation \(n'^{D}\), and validation \(n'^{V}\) for a bottom-level module. If the LLM considers the current design to be simple enough, it decides to invoke \textit{Backtrack Executor}, which we will discuss next.

\minisection{Backtracking}
The invocation of the \textit{Backtrack Executor} can occur not only after the failure of leaf node generation mentioned above but also when all the leaf nodes of any subtree pass self-validation through the \textit{Node Evaluator}. At this point, VeriBToT begins backtracking towards the root node. For each non-leaf node encountered during this process, the \textit{Node Rethinker} is called based on the submodules of this node that have already passed validation, generating new \([n^D, n^V]\), followed by self-validation. During this process, if any non-leaf node fails validation again, backtracking occurs at that node.
At this moment, two choices arise. If the LLM determines that there are issues with the submodule divisions of the current node, it uses a ``removing prompt'' to clear all sub-nodes in the subtree rooted at this node and restarts branching from the backtracked node. Conversely, if the LLM believes that the design of the current node itself is sound but that the task decomposition of the parent node's higher-level module is erroneous, it will continue backtracking to the parent node, removing the backtracked node and its sibling nodes. The LLM will then evaluate in the same manner iteratively until it backtracks to a node where the submodule division is valid and recommences branching.

\begin{table*}[tb!]
\caption{The performance of the ChatGPT-4 in generating Verilog for different hard cases under various reasoning paradigms, the various metrics are the same as those in Table~\ref{tab:Deepseek}.}
    \label{tab:ChatGPT-4}
    \centering
    \footnotesize
    \setlength{\tabcolsep}{5pt} 
    \resizebox{0.88\linewidth}{!}{
    \begin{tabular}{|p{2.5cm}|p{0.6cm}|p{0.6cm}|p{0.6cm}|p{0.6cm}|p{0.6cm}|p{0.6cm}|p{0.6cm}|p{0.6cm}|p{0.6cm}|p{0.6cm}|p{0.6cm}|p{0.6cm}|p{0.6cm}|p{0.6cm}|p{0.6cm}|}
        \hline
        \multirow{2}{*}{Design} & \multicolumn{3}{c|}{IO} & \multicolumn{3}{c|}{CoT} & \multicolumn{3}{c|}{CoT-SC} & \multicolumn{3}{c|}{ToT} & \multicolumn{3}{c|}{VeriBToT}\\
        \cline{2-16}
        & Syn. & Func. & Tok. &  Syn. & Func. & Tok. &  Syn. & Func. & Tok. & Syn. & Func. & Tok.&  Syn. & Func. & Tok. \\
        \hline
        adder\_16bit & \cellcolor{lightGreen}4 & \cellcolor{mycolor3!60}3 & 0.85 & \cellcolor{lightGreen}4 & \cellcolor{lightGreen}4 & 1.88 & \cellcolor{mycolor1}5 & \cellcolor{mycolor1}5 & 3.04 & \cellcolor{lightGreen}4 & \cellcolor{lightGreen}4 & 2.73 & \cellcolor{mycolor1}\textbf{5} & \cellcolor{mycolor1}\textbf{5} & 1.76\\
        \hline
        adder\_32bit & \cellcolor{mycolor1}5 & \cellcolor{mycolor3!60}3 & 1.37 & \cellcolor{mycolor3!60}3 & \cellcolor{mycolor3!60}3 & 1.93 & \cellcolor{mycolor3!60}3 & \cellcolor{mycolor3!60}3 & 2.26 & \cellcolor{mycolor32!60}2 & \cellcolor{mycolor5}0 & 2.19 & \cellcolor{mycolor1}\textbf{5} & \cellcolor{mycolor32!60}2 & 2.01\\
        \hline
        adder\_pipe\_64bit & \cellcolor{mycolor1}5 & \cellcolor{mycolor4}1 & 1.07 & \cellcolor{mycolor3!60}3 & \cellcolor{mycolor5}0 & 1.80 & \cellcolor{mycolor1}5 & \cellcolor{mycolor4}1 & 2.62 & \cellcolor{lightGreen}4 & \cellcolor{mycolor5}0 & 2.65 & \cellcolor{mycolor1}\textbf{5} & \cellcolor{lightGreen}\textbf{4} & 2.37\\
        \hline
        multi\_pipe\_8bit & \cellcolor{mycolor1}5 & \cellcolor{mycolor5}0 & 1.97 & \cellcolor{mycolor3!60}3 & \cellcolor{mycolor5}0 & 1.37 & \cellcolor{mycolor1}5 & \cellcolor{mycolor5}0 & 2.51 & \cellcolor{mycolor32!60}2 & \cellcolor{mycolor5}0 & 2.58 & \cellcolor{lightGreen}4 & \cellcolor{mycolor5}0 & 0.57\\
        \hline
        barrel\_shifter & \cellcolor{mycolor4}1 & \cellcolor{mycolor5}0 & 0.74 & \cellcolor{mycolor1}5 & \cellcolor{mycolor5}0 & 1.2 & \cellcolor{mycolor1}5 & \cellcolor{mycolor5}0 & 1.54 & \cellcolor{mycolor1}5 & \cellcolor{mycolor5}0 & 2.3 & \cellcolor{mycolor1}\textbf{5} & \cellcolor{mycolor5}0 & 2.62\\
        \hline
        width\_8to16 & \cellcolor{mycolor1}5 & \cellcolor{mycolor3!60}3 & 0.85 & \cellcolor{mycolor1}5 & \cellcolor{lightGreen}4 & 1.45 & \cellcolor{lightGreen}4 & \cellcolor{mycolor3!60}3 & 1.61 & \cellcolor{mycolor1}5 & \cellcolor{mycolor3!60}3 & 1.78 & \cellcolor{mycolor1}\textbf{5} & \cellcolor{mycolor3!60}3 & 2.32\\
        \hline
        freq\_div & \cellcolor{mycolor1}5 & \cellcolor{mycolor4}1 & 0.89 & \cellcolor{mycolor1}5 & \cellcolor{lightGreen}4 & 1.70 & \cellcolor{mycolor1}5 & \cellcolor{mycolor3!60}3 & 3.12 & \cellcolor{mycolor1}5 & \cellcolor{lightGreen}4 & 2.05 & \cellcolor{mycolor1}\textbf{5} & \cellcolor{mycolor1}\textbf{5} & 1.99\\
        \hline
        sequence\_detector & \cellcolor{mycolor5}0 & \cellcolor{mycolor5}0 & 1.03 & \cellcolor{mycolor5}0 & \cellcolor{mycolor5}0 & 1.67 & \cellcolor{mycolor4}1 & \cellcolor{mycolor5}0 & 2.13 & \cellcolor{mycolor4}1 & \cellcolor{mycolor5}0 & 2.93 & \cellcolor{mycolor1}\textbf{5} & \cellcolor{mycolor32!60}\textbf{2} & 1.84\\
        \hline   
        dff8r      & \cellcolor{mycolor1}5 & \cellcolor{mycolor5}0 & 0.43 & \cellcolor{mycolor5}0 & \cellcolor{mycolor5}0 & 0.90 & \cellcolor{mycolor5}0 & \cellcolor{mycolor5}0 & 1.46 & \cellcolor{mycolor5}0 & \cellcolor{mycolor5}0 & 1.38 & \cellcolor{mycolor3!60}3 & \cellcolor{mycolor3!60}3& 1.65\\
        \hline
        fsm2s      & \cellcolor{mycolor5}0 & \cellcolor{mycolor5}0 & 0.70 & \cellcolor{lightGreen}4 & \cellcolor{mycolor32!60}2 & 1.00 & \cellcolor{mycolor4}1 & \cellcolor{mycolor4}1 & 1.75 & \cellcolor{mycolor5}0 & \cellcolor{mycolor5}0 & 1.52 & \cellcolor{lightGreen}4 & \cellcolor{mycolor3!60}3& 1.77\\
        \hline
        fsm3comb   & \cellcolor{mycolor5}0 & \cellcolor{mycolor5}0 & 0.66 & \cellcolor{mycolor5}0 & \cellcolor{mycolor5}0 & 0.98 & \cellcolor{mycolor5}0 & \cellcolor{mycolor5}0 & 1.71 & \cellcolor{mycolor3!60}3 & \cellcolor{mycolor3!60}3 & 1.81 & \cellcolor{mycolor1}5 & \cellcolor{mycolor1}5& 1.95\\
        \hline
        gates4     & \cellcolor{mycolor1}5 & \cellcolor{mycolor3!60}3 & 0.42 & \cellcolor{mycolor1}5 & \cellcolor{mycolor1}5 & 0.80 & \cellcolor{mycolor1}5 & \cellcolor{mycolor1}5 & 1.42 & \cellcolor{mycolor3!60}3 & \cellcolor{mycolor3!60}3 & 1.33 & \cellcolor{mycolor1}5 & \cellcolor{mycolor1}5& 1.82\\
        \hline
        popcount3  & \cellcolor{mycolor1}5 & \cellcolor{mycolor5}0 & 0.45 & \cellcolor{mycolor3!60}3 & \cellcolor{mycolor5}0 & 0.70 & \cellcolor{lightGreen}4 & \cellcolor{mycolor4}1 & 1.27 & \cellcolor{mycolor4}1 & \cellcolor{mycolor4}1 & 1.40 & \cellcolor{lightGreen}4 & \cellcolor{mycolor32!60}2& 1.75\\
        \hline
        lemmings1	&\cellcolor{mycolor5}0	&\cellcolor{mycolor5}0& 0.88&	\cellcolor{mycolor32!60}2&	\cellcolor{mycolor32!60}2& 1.38&	\cellcolor{mycolor5}0&	\cellcolor{mycolor5}0& 1.85&	\cellcolor{mycolor5}0&	\cellcolor{mycolor5}0&1.51&	\cellcolor{mycolor1}5&	\cellcolor{mycolor3!60}3&1.77 \\
        \hline
        rule90	    &\cellcolor{mycolor3!60}3	&\cellcolor{mycolor3!60}3&0.76&	\cellcolor{mycolor32!60}2&	\cellcolor{mycolor4}1& 1.09 &	\cellcolor{mycolor32!60}2&	\cellcolor{mycolor32!60}2& 1.54 &	\cellcolor{mycolor3!60}3&	\cellcolor{mycolor4}1& 1.49 &	\cellcolor{mycolor1}5&	\cellcolor{mycolor1}5& 1.65\\
        \hline
        vector3	    &\cellcolor{mycolor3!60}3	&\cellcolor{mycolor4}1&0.50&	\cellcolor{mycolor4}1&	\cellcolor{mycolor4}1&0.95 &	\cellcolor{lightGreen}4&	\cellcolor{mycolor32!60}2& 1.65&	\cellcolor{mycolor32!60}2&	\cellcolor{mycolor5}0& 1.70 &	\cellcolor{mycolor1}5&	\cellcolor{mycolor3!60}3& 1.75\\
        \hline
    \end{tabular}
    }
    
\end{table*}

\minisection{Tree Finality}
When the evaluator believes that the entire RTL design has been correctly completed, the~\textit{Code Aggregator} operation is executed to return the final RTL Verilog output \( y\), and EDA tools are invoked to verify whether it passes the top-level testbench \(x^V \).

Fig.~\ref{fig2} illustrates how a practical RTL design example is developed using the VeriBToT framework. For a 64-bit adder, referred to as \emph{adder\_pipe\_64bit}, the LLM initially cannot complete a correct design within a single Verilog module. As a result, it decomposes this into two submodules: the control module \emph{control\_logic} and the multi-stage adder module \emph{multi\_stage\_adder}. After both submodules are designed and validated through self-validation, the LLM backtracks to the root node and recognizes that they cannot produce a functional adder. It then backtracks further to redefine the submodules, introducing a control module \emph{controller} and a new adder submodule \emph{ripple\_carry\_adder}. Using the testbench, the LLM verifies that the revised design is correct, after which the final RTL code is collected and returned.

\section{Experiments}
\label{sec:experiment}

Our proposed method is compared with standard IO prompting and several common CoT prompting paradigms. To better illustrate the versatility of our approach, we consider four models. Two of them are general-purpose commercial LLMs with strong code generation capabilities: ChatGPT-4 and DeepSeek-Coder-V2~\cite{zhu2024deepseek}. The other two are domain-specific open-source LLMs fine-tuned for Verilog generation: RTLCoder~\cite{RTLCoder} and Thakur~\cite{thakur2022benchmarkinglargelanguagemodels}.~\footnote{Experiments were conducted between March 1–20, 2025.}
We utilized two NL2V (natural language to Verilog code) benchmarks, RTLLM~\cite{lu2024rtllm} and VerilogEval-Human~\cite{liu2023verilogeval}, for our experiments. VerilogEval-Machine, due to its focus on circuit component connections and truth tables, is not considered as it lacks the flexibility for module decomposition. The key performance indicators used for performance measurement are syntactic correctness, functional correctness, the number of tokens consumed, first pass rate (\emph{Pass@1}) five pass rate (\emph{Pass@5}), and the number of successful attempts in the five (\emph{\#Pass@5}).

\subsection{Full Benchmark Experimental Results}

Table~\ref{tab:full} presents the experimental results for \emph{pass@1} and \emph{pass@5} across the entire dataset of two common academic benchmarks. The results indicate that our method effectively enhances model performance, both with the state-of-the-art model ChatGPT-4 and the open-source model Deepseek. 
We did not conduct experiments with different inference paradigms for the domain-specific fine-tuned models (Thakur and RTLCoder), as we observed a significant decline in their general instruction-following ability after fine-tuning. These models were unable to comprehend and adhere to the reasoning paradigms for generation.

\subsection{Ablation Study}

We removed the prompt for generating testbenches and self-validation based on them from the original VeriBToT, leaving the LLM to reflect only on the Verilog generated for each node and determine whether backtracking is necessary, labeled as VeriBToT- in Table~\ref{tab:full}. The experimental results show a significant performance drop, highlighting the effectiveness of the self-validation backtracking mechanism we proposed. ToT can be seen as an ablation of VeriBToT-, where the backtracking mechanism is removed and the model is forced to perform reasoning up to a predetermined depth. As another ablation experiment, the results show that ToT performs worse than VeriBToT-, proving the effectiveness of the backtracking mechanism. This also partially demonstrates that excessive reasoning in LLMs may not necessarily improve performance~\cite{wu2024reasoningrecitingexploringcapabilities}, particularly when CoTs leads to performance degradation compared to IO in the case of RTLLM.

\subsection{Hard Case Experimental Results}

We conducted a detailed case analysis on the function and syntax pass rates (\emph{\#pass@5}) described in the RTLCoder paper~\cite{RTLCoder}, specifically focusing on the relatively low performance in hard cases. Hard cases are defined as those circuits for which the direct IO prompt failed in the functional \emph{pass@1} experiment of Table~\ref{tab:full}. Due to space limitations, we only present 8 cases each for RTLLM and VerilogEval-Human. The results, as shown in \Cref{tab:Deepseek,tab:ChatGPT-4}, demonstrate that our method encourages deeper model reasoning, leading to improvements in both syntactic and functional correctness compared to IO and other CoT modes. Moreover, token consumption does not increase significantly; in some instances, it is even lower than that of ToT and CoT-SC.

\definecolor{mycolor1}{rgb}{0.4, 0.8, 0.1}
\definecolor{mycolor3}{rgb}{0.8, 0.5, 0}
\definecolor{mycolor32}{rgb}{1, 0.8, 0.1}
\definecolor{mycolor4}{rgb}{1, 0.75, 0.7}
\definecolor{mycolor5}{rgb}{0.9, 0.3, 0.3}

\subsection{Token Consumption Evaluation}
\begin{figure}[tb]
    \center
    \includegraphics[width=9cm]{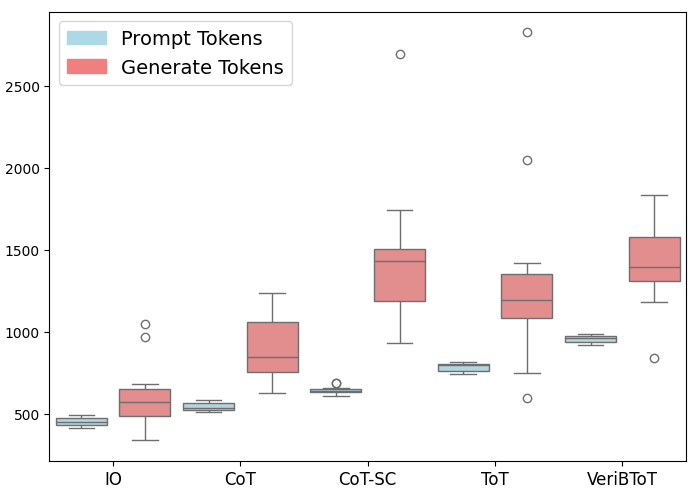}
    \caption{
        The token consumption comparison between different CoT paradigms for RTL generation.
    }
    \label{fig:token}
\end{figure}

To assess VeriBToT’s token overhead, we analyzed token usage across reasoning paradigms, as shown in Fig.~\ref{fig:token}. The results indicate that although VeriBToT increases prompt token consumption, the tokens generated in context remain comparable to those of CoT-SC and ToT. Given the additional context introduced—such as testbench code and natural language sub-module descriptions—the overall increase is relatively small. This demonstrates the efficiency of the backtracking mechanism in controlling token usage.

\section{Conclusion}
\label{sec:conclu}

In this paper, we present VeriBToT, a novel CoT architecture for automating RTL design with LLMs. Inspired by top-down IC design and Design for Verification (DFV), VeriBToT enables self-decoupling and self-verification of thought nodes, represented through a BackTrack Tree. This improves the model’s understanding of Verilog design workflows, and our experiments show the broad applicability of the approach. We believe VeriBToT can serve as a unified reasoning paradigm for large-scale circuit generation with LLMs.

\newpage

\bibliographystyle{IEEEtran}
\bibliography{reference}
\end{document}